\documentclass[reprint,superscriptaddress,amsmath,amssymb,nofootinbib,aps,twocolumn]{revtex4-1}

\usepackage{graphicx}
\usepackage{amsmath}
\usepackage{multirow}
\usepackage[colorlinks=true,citecolor=blue,linkcolor=blue,urlcolor=blue]{hyperref}

\makeatletter
\renewcommand{\frontmatter@abstractheading}{\textbf{Abstract:}}
\makeatother

\begin{document}

\title{Constraints on Compact Dark Matters Population from microlensing Effect of Gravitational Wave for the third-generation gravitational Wave Detector}

\author{Xin-Yi Lin}
\affiliation{School of Physics and Astronomy, Beijing Normal University, Beijing 100875, China}

\author{Xi-Jing Wang}
\affiliation{Department of Astronomy, School of Physics and Technology, Wuhan University, Wuhan 430072, China}

\author{Huan Zhou}
\email{\textcolor{black}{Corresponding author:} huanzhou@yangtzeu.edu.cn}
\affiliation{School of Physics and Optoelectronic Engineering, Yangtze University, Jingzhou, 434023, China}
\affiliation{Department of Astronomy, School of Physics and Technology, Wuhan University, Wuhan 430072, China}

\author{Zhengxiang Li}
\email{\textcolor{black}{Corresponding author:} zxli918@bnu.edu.cn}
\affiliation{Department of Astronomy, Beijing Normal University, Beijing 100875, China}
\affiliation{Institute for Frontiers in Astronomy and Astrophysics, Beijing Normal University, Beijing 102206, China}

\begin{abstract}
Since the first direct detection of gravitational waves (GWs) from a binary black-hole merger by the LIGO-Virgo Collaboration, GWs have emerged as a powerful probe of astrophysics and cosmology. If extragalactic compact dark matter (DM) candidates, such as primordial black holes, constitute a non-negligible fraction of the DM component over a broad mass range, they can imprint distinctive microlensing signatures on GW signals. In this work, we develop a framework to infer the population properties of extragalactic compact DM by simulating microlensing GWs in the frequency band of the third-generation ground-based GW detector, the Einstein Telescope, and analyzing them via hierarchical Bayesian inference. For a population described by a power-law mass function, we show that the detection of only several microlensing GW events among $10^4$ binary black-hole coalescences can constrain the compact DMs abundance to the level of $\sim 10^{-3}$. 
This forecast demonstrates that searching for and identifying microlensing signatures in future GW observations can provide a complementary approach to constraining compact DM scenarios.

\vspace{0.6em}
\end{abstract}

\maketitle

\section{Introduction}\label{sec1}
The cosmic abundance of dark matter (DM) has been measured with high precision by cosmological observations, in particular through analyses of the cosmic microwave background (CMB) anisotropies, and is found to account for approximately $26\%$ of the total energy density of the Universe~\cite{Planck2018}. Despite such robust macroscopic evidence, the nature of dark matter remains unknown~\cite{Hui:2021tkt,OHare:2024nmr}. A variety of DM scenarios predict the existence of compact objects, including primordial black holes (PBHs)~\cite{Hawking:1971ei,Carr:1974nx,Carr:1975qj}, mini-halos~\cite{Ricotti:2009bs,Hardy:2016mns}, boson stars~\cite{Kaup:1968zz,Brito:2015pxa}, and fermion stars~\cite{Kouvaris:2015rea}. Among these candidates, PBHs stand out as a long-standing and compelling possibility for DM, and have drawn extensive attention across astrophysics and cosmology (see the recent reviews~\cite{Sasaki2018,Green2020,Carr2020,Carr:2026hot}). Gravitational lensing offers a powerful tool to probe compact DM across a broad mass range~\cite{Liao2022}, through four main observational signatures: (1) microlensing-induced stellar luminosity variations in the Milky Way and nearby galaxies~\cite{Niikura:2017zjd,Mroz:2024mse}; (2) time-delayed multi-peaked structures in transient sources such as fast radio bursts (FRBs) and gamma-ray bursts (GRBs)~\cite{Munoz:2016tmg,Liao:2020wae,Nemiroff:2001bp,Ji:2018rvg}; (3) waveform distortions in both gravitational waves (GWs) and FRBs due to microlensing effect~\cite{Jung2019,LIGOScientific:2025cwb,CHIMEFRB:2022xzl}; and (4) angularly separated multiple images in persistent sources like compact radio sources~\cite{Wilkinson:2001vv,Zhou:2021tvp}.

On September 14, 2015, the Advanced LIGO detectors~\cite{LIGOScientific:2014pky,LIGOScientific:2016emj}, together with Virgo~\cite{VIRGO:2014yos}, made the first direct detection of a GW signal from a binary black-hole (BBH) merger, GW150914~\cite{LIGO2016}. 
In 2019, the baseline installation of KAGRA was completed~\cite{KAGRA:2020tym,Somiya:2011np,Aso:2013eba}. Since then, the LIGO-Virgo-KAGRA network, commonly referred to as the LVK Collaboration, has detected approximately $\mathcal{O}(10^2)$ GW events during the O1--O4 observing runs~\cite{LIGOScientific:2016lio,LIGOScientific:2017zic, LIGOScientific:2019fpa,LIGOScientific:2020tif,Jung2019,LIGOScientific:2025cwb,LIGOScientific:2018mvr,LIGOScientific:2020ibl,LIGOScientific:2021usb,KAGRA:2021vkt,LIGOScientific:2025pvj,LIGOScientific:2026ctl}. The growing number of GW detections has opened a new window onto the Universe, providing unprecedented opportunities to investigate astrophysics and cosmology~\cite{LIGOScientific:2017adf, LIGOScientific:2019zcs,LIGOScientific:2018gmd,LIGOScientific:2021aug,Jana:2022shb,Jana:2024dhc,Zhou:2026uex}. Similar to electromagnetic waves, GWs can be gravitationally lensed by massive objects, such as galaxies and compact DMs. However, unlike stellar light, GWs detected by ground-based observatories have much longer wavelengths, making wave-optics effects important in many lensing scenarios~\cite{Ohanian:1974ys,Nakamura:1997sw,Nakamura:1999uwi,Takahashi:2003ix,Dai:2018enj,Boileau:2020rpg,Gao:2021sxw,Fairbairn2022,Caliskan:2022hbu,Leung:2023lmq,GilChoi2023,Savastano2023,Cheung2024,Villarrubia-Rojo:2024xcj}. These diffraction effects can distort the GW waveform and produce characteristic interference fringes. Since~\citet{Jung2019} first proposed that GW microlensing observed by Advanced LIGO could be used to probe compact DMs, GW microlensing has been extensively studied as a means of constraining compact-DMs populations~\cite{Liao2020,Basak2021,Urrutia2021,Zhou2022,Wang2021,Urrutia2023,LIGO2023,Barsode2024,Diego2019,Guo2022,Gais:2022xir,Lin2025}.~Many previous studies have focused on constraining the abundance of compact DMs, quantified by the fraction $f_{\rm CO}\equiv\Omega_{\rm CO}/\Omega_{\rm DM}$, with $\Omega_{\rm CO}$ and $\Omega_{\rm DM}$ denoting the present-day density parameters of compact DMs and total DMs, respectively. These studies have mainly relied on null searches for microlensing signatures in current GW observations~\cite{Jung2019,Liao2020,Wang2021,Urrutia2021,Zhou2022,Urrutia2023,Basak2021,LIGO2023,Barsode2024}, while analyses of compact DM abundance based on simulated or identified lensed GW events have also been explored~\cite{Basak2021,Zhou2022,Gais:2022xir}. In particular, Refs.~\cite{Basak2021,Zhou2022} studied compact DM abundance constraints with third-generation GW detectors, considering extended mass functions and a scenario with a positively identified lensed event, but did not account for selection effects or measurement uncertainties. In addition, Ref.~\cite{Gais:2022xir} considered hierarchical inference for a power-law lens mass spectrum to constrain the local compact DM abundance, but did not account for selection effect in the detection process. Building on these studies, we establish a hierarchical Bayesian inference framework that accounts for selection effects and measurement uncertainties for microlensing studies with third-generation GW detectors. We focus on whether an ensemble of individually detected microlensing GW events can simultaneously constrain both the compact DM abundance \(f_{\rm CO}\) and the slope \(\gamma\) of an extended lens mass function. Though there is no conclusive evidence to confirm the existence of microlensing GW signals, with the significantly improved sensitivity of next-generation ground-based GW detectors, e.g., the Einstein Telescope (ET)~\cite{Branchesi:2023mws,Punturo:2010zza} and Cosmic Explorer (CE)~\cite{Reitze:2019iox}, the detections of microlensing GW events are expected to become possible. Such detections would provide valuable information on the population properties of compact DMs.

In this work, we present a proof-of-principle framework that connects the simulation of microlensing GW signals with the extraction of population information for the lensing objects. As a compelling class of compact DMs candidates, we focus on the compact DMs in the mass range of $10 \sim 10^3~M_{\odot}$ as lenses to demonstrate our methodology. Furthermore, taking the Einstein Telescope (ET) as a representative example, we provide quantitative projections for the microlensing detection capabilities of next-generation ground-based GW detectors in the frequency band from $5$ to $10^3\,{\rm Hz}$. In this frequency range, the compact DMs masses are well matched to the Hz-kHz frequency band of ET, where wave optics microlensing effects are expected to be most relevant. The remainder of this paper is organized as follows. In Section~\ref{sec2}, we introduce the theoretical framework for GW microlensing in the wave-optics regime. Section~\ref{sec3} describes the simulation of microlensing events. In Section~\ref{sec4}, we present the hierarchical Bayesian inference (HBI) framework used to extract the population properties of compact DMs, such as its mass function and abundance, from an ensemble of microlensing GW events. The resulting constraints are reported in Section~\ref{sec5}. Finally, we summarize our conclusions in Section~\ref{sec6}. Throughout this paper, we adopt the concordance $\Lambda$CDM cosmology with the best-fitting parameters from the latest Planck observations~\cite{Planck2018}, and use natural units with $G=c=1$ in all equations.

\section{Theoretical framework for microlensing of gravitational wave}
\label{sec2}
To describe the lensing effect on GWs, we consider a lens with gravitational potential $U(\vec{r})$ embedded in an FLRW background. The corresponding perturbed metric can be written as
\begin{equation}\label{eq2-1-1f}
ds^2=-(1+2U(\vec{r}))dt^2+a(t)^2(1-2U(\vec{r}))d\vec{r}^2.
\end{equation}
A GW is treated as a small metric perturbation $h_{\mu\nu}$ propagating on this background metric $g_{\mu\nu}^{(0)}$. The scalar amplitude $h$ associated with the perturbation satisfies the wave equation
\begin{equation}\label{eq2-1-1}
\partial_{\mu}(\sqrt{-g^{(0)}}g^{(0)\mu\nu}\partial_{\nu}h)=0.
\end{equation}
Since lensing can be treated as a local process occurring in the vicinity of the lens, the time dependence of the scale factor can be neglected. 
In Fourier space, the wave equation for $h$ then reduces to the Helmholtz equation
\begin{equation}\label{eq2-1-2}
(\nabla^2+4\pi^2f^2)\bar{h}=16\pi^2f^2U\bar{h},
\end{equation}
where $f$ is the frequency and $\bar{h}$ denotes the Fourier transform of $h$. The dimensionless amplification factor is defined as
\begin{equation}\label{eq2-1-3}
F(f)=\frac{\bar{h}_{\rm L}(f)}{\bar{h}_0(f)},
\end{equation}
where $\bar{h}_{\rm L}(f)$ and $\bar{h}_0(f)$ are the lensed and unlensed $(U=0)$ wave amplitudes, respectively. In the thin-lens and point-source approximations, the amplification factor can be written as~\cite{Ohanian:1974ys,Nakamura:1997sw,Nakamura:1999uwi,Takahashi:2003ix,Dai:2018enj,Boileau:2020rpg,Gao:2021sxw,Fairbairn2022,Caliskan:2022hbu,Leung:2023lmq,GilChoi2023,Savastano2023,Cheung2024,Villarrubia-Rojo:2024xcj}
\begin{equation}\label{eq2-1-4}
F(f)=\frac{D_{\rm S}\xi_0^2(1+z_{\rm L})f}{iD_{\rm LS}D_{\rm L}}\int d^2\vec{x}~\exp[2\pi i ft_{\rm d}(\vec{x},\vec{y})],
\end{equation}
where $\xi_0\equiv\theta_{\rm E}D_{\rm L}$ is a normalization scale set by the Einstein radius $\theta_{\rm E}$, which is the typical scale length of the impact parameter in gravitational lensing. And $z_{\rm L}$ is the lens redshift, $D_{\rm S}$, $D_{\rm L}$, and $D_{\rm LS}$ denote the angular-diameter distances to the source, to the lens, and between the lens and the source, respectively. The dimensionless vectors $\vec{x}$ and $\vec{y}$ are defined as
\begin{equation}\label{eq2-1-4f}
\begin{split}
&\vec{x}\equiv\frac{\vec{\xi}}{\xi_0},\\
&\vec{y}\equiv\frac{D_{\rm L}}{D_{\rm S}\xi_0}\vec{\eta},
\end{split}
\end{equation}
where $\vec{\xi}$ and $\vec{\eta}$ are the physical coordinates in the lens plane and source plane, respectively. The arrival time at the observer from the source position $t_{\rm d}$ is given by
\begin{equation}\label{eq2-1-5}
\begin{split}
&t_d(\vec{x},\vec{y})=\frac{D_{\rm S}\xi_0^2(1+z_{\rm L})}{D_{\rm L}D_{\rm LS}}\times\\
&\bigg[\frac{1}{2}|\vec{x}-\vec{y}|^2-\psi(\vec{x})+\phi_{\rm m}(\vec{y})\bigg],
\end{split}
\end{equation}
where $\psi(\vec{x})$ is the dimensionless deflection potential, and $\phi_{\rm m}(\vec{y})$ is chosen such that the minimum arrival-time delay is zero.

In this work, we take the point mass lens model. For this model, using the rescaled radial coordinates $x=|\vec{x}|$ and $y=|\vec{y}|$, one has
$\psi(\vec{x})=\ln x$ and $\phi_{\rm m}(\vec{y})=(x_{\rm m}-y)^2/2-\ln x_{\rm m}$, with $x_{\rm m}=(y+\sqrt{y^2+4})/2$. In this case, the amplification factor admits the analytic form~\cite{Takahashi:2003ix}
\begin{equation}\label{eq2-1-6}
\begin{split}
F(f)=\exp\bigg[\frac{\pi w}{4}+i\frac{w}{2}\bigg(\ln \frac{w}{2}-2\phi_{\rm m}(y)\bigg)\bigg]\\
\times\Gamma\bigg(1-\frac{iw}{2}\bigg){_{1}F_1}\bigg(i\frac{w}{2},1,\frac{i}{2}w y^2\bigg),
\end{split}
\end{equation}
where $w=8\pi m_{\rm L}(1+z_{\rm L})f$ is the dimensionless frequency, and ${_{1}F_1}$ is the confluent hypergeometric function. Therefore, for a point-mass lens, the amplification factor depends only on the redshifted lens mass $M_{\rm L}^{z}\equiv(1+z_{\rm L})m_{\rm L}$ and the source position $y$. 

The geometrical-optics limit applies when the GW wavelength is much smaller than the Schwarzschild radius of the lens. In this regime, the amplification factor can be well approximated by the discrete contributions from multiple images,
\begin{equation}\label{eq2-1-6f}
\begin{split}
F(f)=|\mu_{+}|^{1/2}-i|\mu_{-}|^{1/2}\exp(i2\pi f\Delta t_{\rm d}),
\end{split}
\end{equation}
where the magnification of each image is $\mu_{\pm} = 1/2 \pm (y^2 + 2)/(2y\sqrt{y^2 + 4})$, and $\Delta t_{\rm d} = 4M_{\rm L}^{z}[y\sqrt{y^2 + 4}/2+\ln((\sqrt{y^2 + 4}+y)/(\sqrt{y^2 + 4}-y)]$ is the time delay between the double images. In the numerical implementation, we adopt this geometrical-optics approximation when $w\geq 10$, since the direct evaluation of Eq.~\eqref{eq2-1-6} becomes computationally inefficient and numerically unstable at large $w$. The largest time delay expected within the compact-DM mass range considered in this work is of order $\mathcal{O}(10^{-2})\,{\rm s}$, implying that the two lensed images from the same source overlap in the time domain rather than being resolved as separate events.

\begin{figure*}
    \centering
    \includegraphics[width=0.48\textwidth, height=0.48\textwidth]{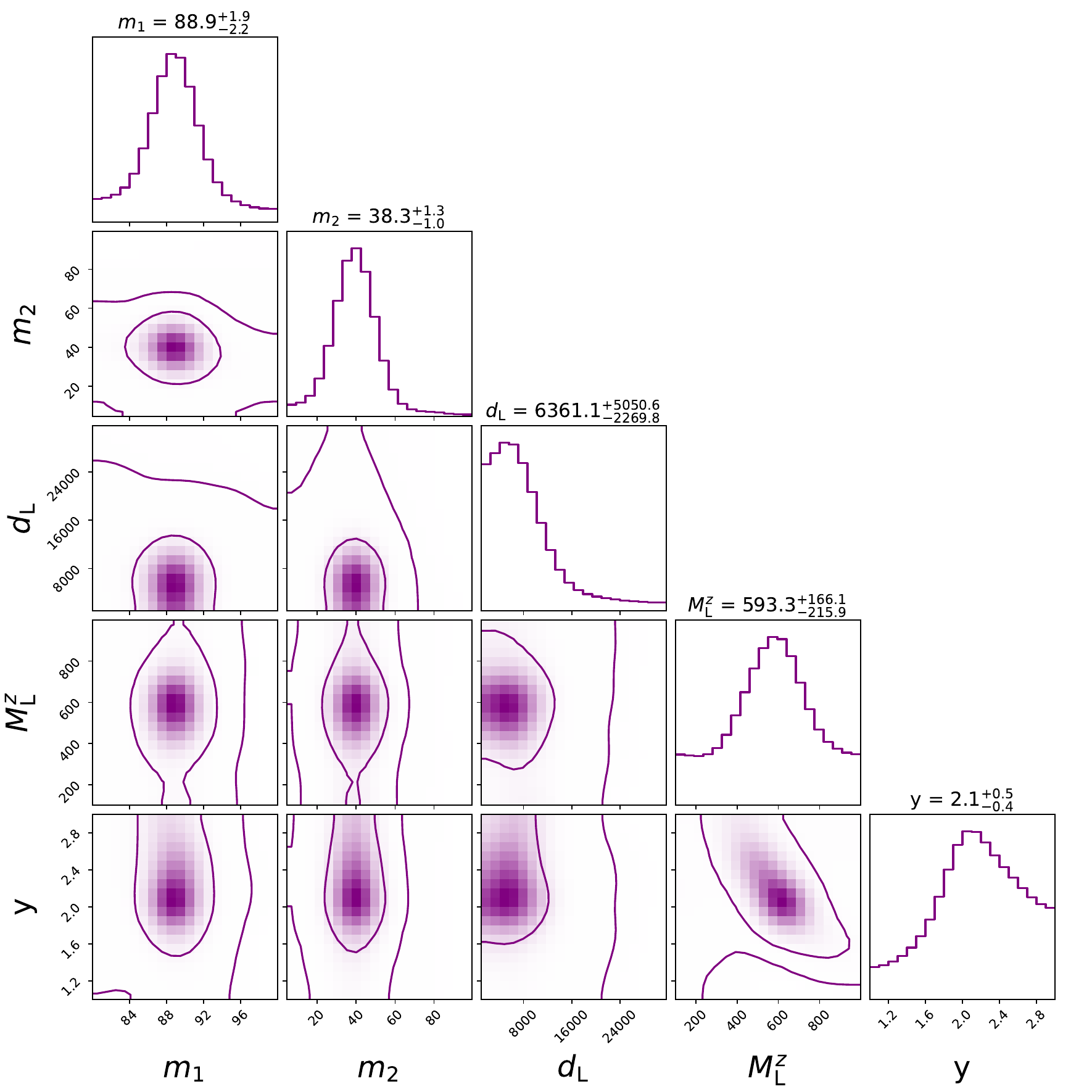}
    \includegraphics[width=0.48\textwidth, height=0.48\textwidth]{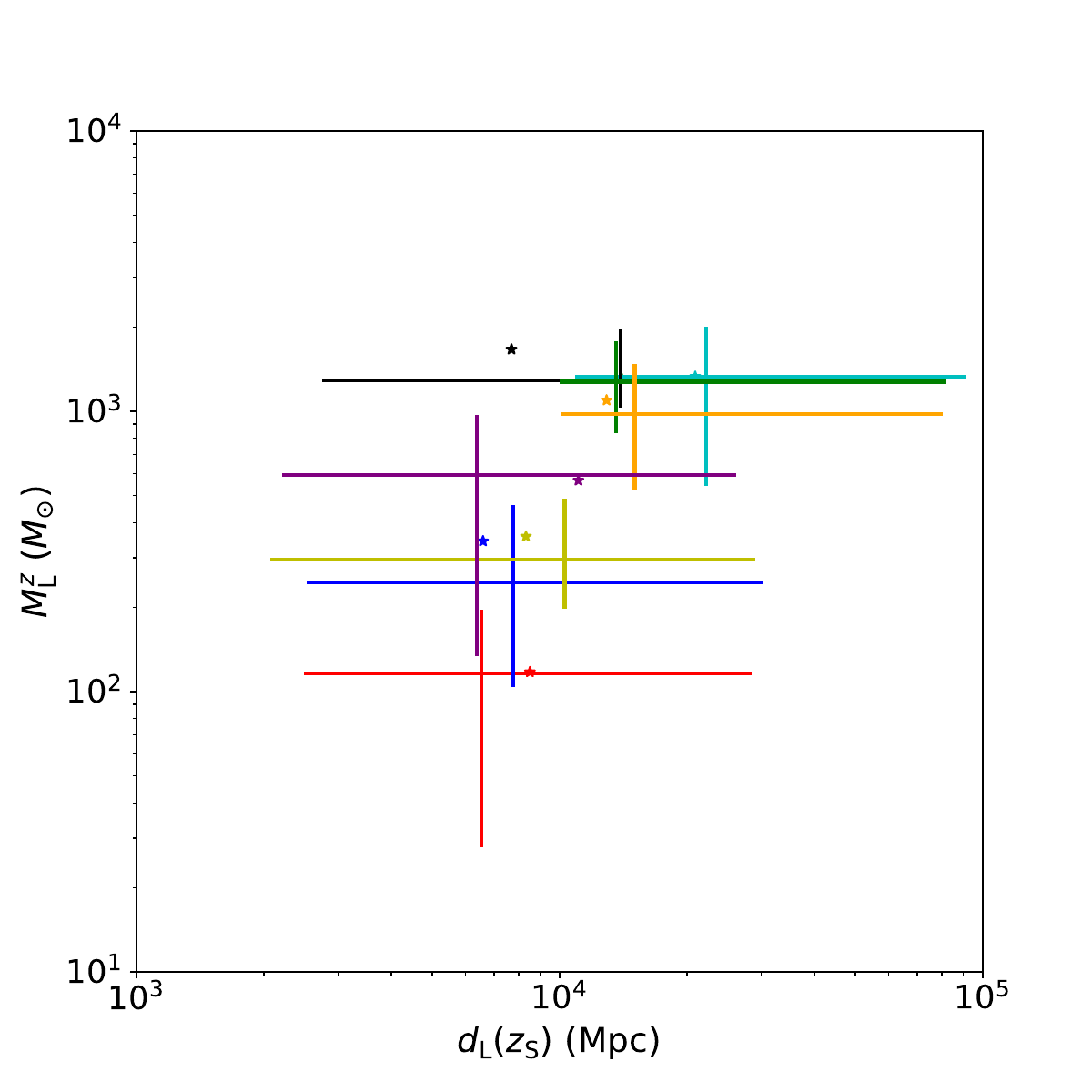}
     \caption{{\bf Left:} Posterior distribution for one representative event among the eight detectable microlensing GW events. {\bf Right:} Comparison between the inferred and injected parameters for the eight detectable microlensing GW events. The crosses indicate the $2\sigma$ credible regions of the source luminosity distance and the redshifted lens mass inferred from the posterior distributions, while the scattered points denote the corresponding injected values.}\label{fig1}
\end{figure*}

\section{Simulation}
\label{sec3}

In this section, we describe the simulation of GW signals and the generation of microlensing GW events. 
To ensure both physical consistency and computational efficiency in the simulation of microlensing GW events, we adopt a five-step procedure:
\begin{itemize}
\item\textbf{Generating the detectable BBH events:} 
Although the astrophysical origin of BBH mergers remains an open question, we adopt BBH coalescences as the GW sources in this study for the following reasons. These systems typically have high signal-to-noise ratios (SNRs), robust waveform templates, and high event rates. Moreover, most GW events observed so far by the LVK Collaboration are associated with BBH mergers. The redshift distribution of the GW sources is drawn from a phenomenological merger-rate model,
\begin{equation}\label{eq3-1-1}
P_{\rm BBH}(z_{\rm S})=\frac{1}{Z_{\rm BBH}}\frac{R_{\rm BBH}(z_{\rm S})}{1+z_{\rm S}}\frac{dV_{\rm c}}{dz_{\rm S}},
\end{equation}
where $Z_{\rm BBH}$ is the normalization constant, and $R_{\rm BBH}$ is the merger rate per unit comoving volume in the source frame. $R_{\rm BBH}$ is given by~\cite{Urrutia2021,Mukherjee2021}
\begin{equation}\label{eq3-1-2f}
\begin{split}
&R_{\rm BBH}(z_{\rm S})=R_0\times\\
&\int_{t_{\min}}^{t_{\max}}R_{\rm f}(t(z_{\rm S})-t_{\rm delay})P(t_{\rm delay}) dt_{\rm delay},
\end{split}
\end{equation}
where $t_{\min}=50~{\rm Myr}$, $t_{\max}$ is set to the Hubble time, and $P(t_{\rm delay})\propto t_{\rm delay}^{-1}$ is the delay-time distribution, and $R_{\rm f}(t(z_{\rm S})-t_{\rm delay})$ follows the redshift dependence of the Star Formation Rate (SFR), which is parameterized as~\cite{Madau2016,Urrutia2021}
\begin{equation}\label{eq3-1-2}
\begin{split}
R_{\rm f}(z_{\rm S})=[1+(1+z_{\rm p})^{-\lambda-\kappa}]\times\\
\frac{(1+z_{\rm S})^{\lambda}}{1+\bigg(\frac{1+z_{\rm S}}{1+z_{\rm p}}\bigg)^{\lambda+\kappa}}.
\end{split}
\end{equation}
For $z_{\rm p}\gg 0$ and $\lambda+\kappa>0$, this model scales approximately as $(1+z)^{\lambda}$ at low redshift and as $(1+z)^{-\kappa}$ at high redshift. In Eq.~\eqref{eq3-1-2}, we adopt $\lambda=2.6$, $\kappa=3.6$, and $z_{\rm p}=2.2$~\cite{Madau2016}. For the mass distribution of BBHs, we take a power-law distribution for the heavier black hole~\cite{Liao2020,Basak2021,Urrutia2021,Zhou2022},
\begin{equation}\label{eq3-1-3}
\begin{split}
&P_{\rm BBH}(m_{\rm s})=Z_{\rm m}m_{\rm s}^{-\alpha}\times\\
&\mathcal{H}(m_{\rm s}-m_{\min})\mathcal{H}(m_{\max}-m_{\rm s}),
\end{split}
\end{equation}
where $\mathcal{H}$ is the Heaviside step function and $Z_{\rm m}$ is the normalization constant,
\begin{equation}\label{eq3-1-4}
Z_{\rm m}=\left\{
\begin{aligned}
\frac{1-\alpha}{M_{\max}^{1-\alpha}-M_{\min}^{1-\alpha}},\quad\alpha\neq1,\\
\frac{1}{\ln(M_{\max}/M_{\min})},\quad\alpha=1.
\end{aligned}
\right.
\end{equation}

The primary mass $m_1$ follows Eq.~\eqref{eq3-1-3} with $\alpha=2.35$ over the range $m_1\in[5,100]\,M_{\odot}$, while the secondary mass $m_2$ is drawn uniformly from the interval $[m_1/18,m_1]$. For the remaining source parameters, we fix the component spins to $s_1=s_2=0$ and set the orbital eccentricity to $e=0$. The orbital inclination $\cos\iota\in\mathcal{U}[-1,1]$, coalescence phase $\phi_{\rm c}\in\mathcal{U}[0,2\pi]$, sky-location angles $\alpha\in\mathcal{U}[0,2\pi]$ and $\sin\delta\in\mathcal{U}[-1,1]$, and polarization angle $\psi\in\mathcal{U}[0,\pi]$ are drawn from uniform distributions. We use the Python package \textsf{PyCBC} to generate frequency-domain~\textsf{IMRPhenomD} waveforms for the unlensed signals $\bar{h}_0(f)$~\cite{alex_nitz_2024_10473621}.

Based on the redshift and mass distributions described above, we simulate $10^4$ ET-detectable unlensed BBH events with ${\rm SNR}\geq 8$. The SNR of an unlensed GW signal is computed as
\begin{equation}\label{eq3-3-1}
{\rm{SNR}_0}=\sqrt{\left\langle \bar{h}_0|\bar{h}_0\right\rangle}=\sqrt{4\int\frac{|\bar{h}_0(f)|^2}{S_{\rm n}(f)}df},
\end{equation}
where $S_{\rm n}(f)$ is the power spectral density for ET\cite{ETSensitivities}, and the inner product between two waveforms $\bar{h}_1$ and $\bar{h}_2$ is given by
\begin{equation}
\begin{split}
&\left\langle \bar{h}_1 \middle| \bar{h}_2 \right\rangle = 2\times\\
&\int_{f_{\min}}^{f_{\max}}
\frac{\bar{h}_1^{*}(f)\bar{h}_2(f) +
\bar{h}_2^{*}(f)\bar{h}_1(f)}{S_n(f)}\,df ,
\end{split}
\end{equation}
where the frequency limits $f_{\min}$ and $f_{\max}$ are determined by the waveform segment used in our analysis, which starts from $10^3~{\rm s}$ before coalescence and includes the merger and ringdown phases. The quoted SNR is calculated using the response of a single ET interferometer, E1.

\item\textbf{Generating microlensing GW events:} 
Assuming that compact DMs are homogeneously distributed in comoving volume, the probability that a GW source at redshift $z_{\rm S}$ is lensed by an intervening compact DM is given by the optical depth\cite{Zhou:2022yeo,Urrutia2021},
\begin{equation}\label{eq3-3-2}
P_{\rm L}(z_{\rm S})=1-\exp(-\tau(z_{\rm S})).
\end{equation}
For each BBH GW signal, we draw a random number uniformly from the interval $[0,1]$ and identify the event as lensed if this number is smaller than $P_{\rm L}(z_{\rm S})$. The optical depth $\tau(z_{\rm S})$ is written as\cite{Wang2021}
\begin{equation}\label{eq3-3-3}
\begin{split}
\tau(z_{\rm S})=&\int_0^{+\infty}dm_{\rm L}\int_0^{z_{\rm S}}dz_{\rm L}\int_0^{y_0}dy\\
&\tau(m_{\rm L},z_{\rm L},y),
\end{split}
\end{equation}
where $y_0$ is the maximum impact parameter. Here, $\tau(m_{\rm L},z_{\rm L},y)$ denotes the differential optical depth for a lens with mass $m_{\rm L}$, lens redshift $z_{\rm L}$, and impact parameter $y$, and is given by\cite{Wang2021}
\begin{equation}\label{eq3-3-4}
\begin{split}
\tau(m_{\rm L},z_{\rm L}, y)=&3f_{\rm CO}\Omega_{\rm DM}y\psi(m_{\rm L}, \boldsymbol p_{\rm mf})\times\\
&\frac{H_0^2(1+z_{\rm L})^2}{H(z_{\rm L})}\frac{D_{\rm L}D_{\rm LS}}{D_{\rm S}},
\end{split}
\end{equation}
where $\boldsymbol p_{\rm mf}$ denotes the set of hyperparameters that characterize the compact DM mass function. We set the injected value $f_{\rm CO}=10^{-2.5}$. 

\item\textbf{Generating the parameters of lens:} When a GW event is identified as lensed, the lens redshift $z_{\rm L}$ is drawn from the conditional probability distribution implied by the differential optical depth in Eq.~\eqref{eq3-3-4},
\begin{equation}\label{eq3-3-5}
P(z_{\rm L})=\frac{\tau(m_{\rm L},z_{\rm L},y)}{\int_0^{z_{\rm S}}dz_{\rm L}~\tau(m_{\rm L},z_{\rm L},y)}.
\end{equation}
Similarly, the distribution of the impact parameter $y$ can be obtained from Eq.~\eqref{eq3-3-4} over the range $y\in[0,y_0]$,
\begin{equation}\label{eq3-3-6}
P(y)=\frac{\tau(m_{\rm L},z_{\rm L},y)}{\int_0^{y_0}dy~\tau(m_{\rm L},z_{\rm L},y)}=\frac{2}{y_0^2}y.
\end{equation}
As stated above, we set $y_0=5$. For the extended lens mass distribution $\psi(m_{\rm L},\boldsymbol p_{\rm mf})$, defined in the same way as in Eq.~\eqref{eq2-2-4}, we adopt a power-law mass function with a minimum mass $M_{\min}=10\,M_{\odot}$ and a maximum mass $M_{\max}=10^3\,M_{\odot}$. For the mass function adopted in this work with the range of $M$ fixed, the lens mass distribution is then given by
\begin{equation}\label{eq3-2-1}
\begin{split}
\psi(m_{\rm L},\gamma)=&\mathcal{N}_{\rm pl}m_{\rm L}^{\gamma-1}
\mathcal{H}(m_{\rm L}-M_{\min})\times\\
&\mathcal{H}(M_{\max}-m_{\rm L}),
\end{split}
\end{equation}
where $\mathcal{N}_{\rm pl}$ is the normalization constant. In these simulations, we set the power-law index to $\gamma=1$. This choice corresponds to a flat mass distribution per unit lens mass over the mass range of compact DMs in our work.

\item\textbf{Computing lensed criterion:} After generating the unlensed and lensed GW waveforms using the simulated source and lens parameters, we select detectable microlensing events based on the mismatch $\epsilon$ and the lensed signal-to-noise ratio ${\rm SNR}_{\rm L}$, following Ref.~\cite{Savastano2023}. The selection criterion is given by
\begin{equation}\label{eq3-3-9}
\epsilon\times {\rm {SNR_{L}}}^2>1,
\end{equation}
where the mismatch $\epsilon$ between the lensed and unlensed waveforms is defined as
\begin{equation}\label{eq3-3-10}
\epsilon=1-\frac{\left\langle \bar{h}_{\rm L}|\bar{h}_0\right\rangle}{\sqrt{\left\langle \bar{h}_{\rm L}|\bar{h}_{\rm L}\right\rangle\left\langle \bar{h}_0|\bar{h}_0\right\rangle}}.
\end{equation}

\item\textbf{Extract parameters of microlensing GWs:}
We use Bayesian inference to obtain the posterior distributions of the parameters of interest,
$[m_1,m_2,d_L(z_S),M_{\rm L}^z,y]$, while simultaneously sampling and marginalizing over the nuisance parameters, including the orbital inclination $\iota$, coalescence phase $\phi_c$, \textbf{sky-location angles $(\alpha,\delta)$}, and polarization angle $\psi$. The likelihood for each GW signal can be expressed as an inner product~\cite{Lin2025}
\begin{equation}\label{eq3-3-11}
p(d|\theta)\propto\exp\bigg[-\frac{1}{2}\left\langle d-\bar{h}(\theta)|d-\bar{h}(\theta)\right\rangle\bigg],
\end{equation}
where $\bar{h}(\theta)$ represents the model of microlensing GW waveform. In the parameter estimation, we sample all source and lens parameters together with the nuisance parameters mentioned above. The posterior distributions are obtained using the dynamic nested sampler implemented in \texttt{dynesty} with the random-walk sampling method. We adopt zero-noise injections, i.e., the simulated data contain the lensed waveform without additional detector noise realizations.

\end{itemize}

\begin{figure*}
    \centering
    \includegraphics[width=0.48\textwidth, height=0.36\textwidth]{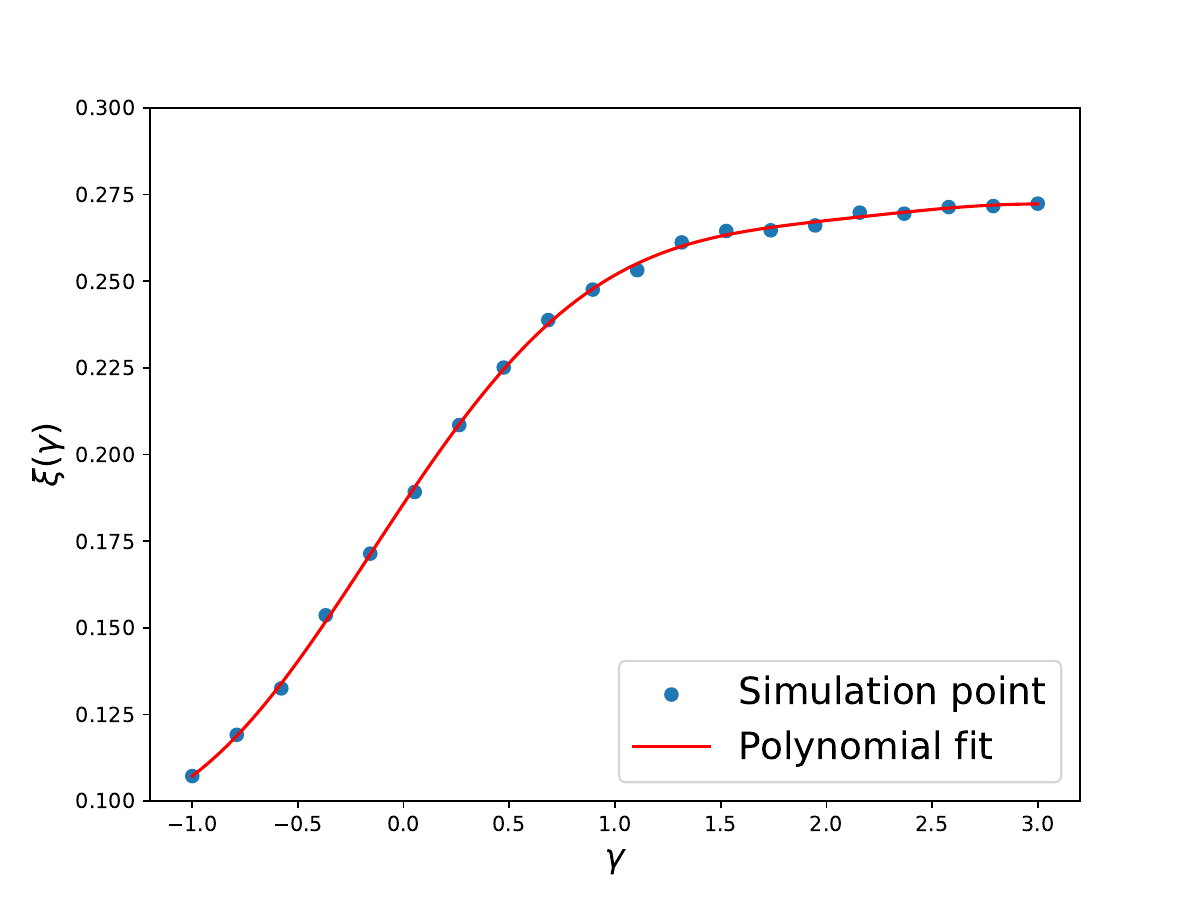}
     \includegraphics[width=0.48\textwidth, height=0.36\textwidth]{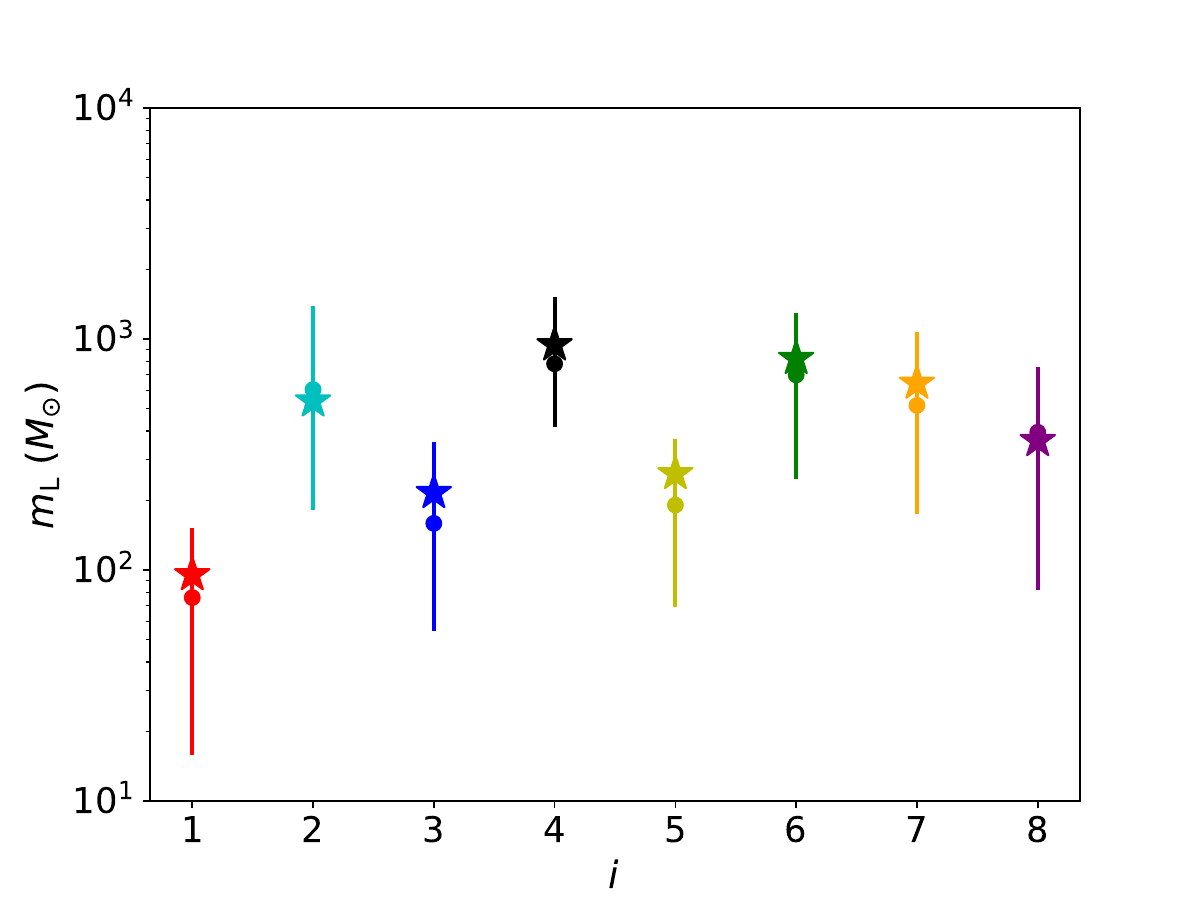}
     \caption{{\bf Left:} Detectable fraction of microlensing GW events, as defined in Eq.~\eqref{eq2-2-8f}. The blue points represent the simulated results, while the solid red curve shows the corresponding polynomial fit. {\bf Right:} Inferred lens masses for the eight detectable microlensing GW events. The solid lines with points indicate the median values and $95\%$ credible intervals obtained from the posterior distributions of the redshifted lens mass, $p(M_{\rm L}^{z}|d_i)$, and the GW-source luminosity distance, $p(d_{\rm L}(z_{\rm S})|d_i)$, shown in Fig.~\ref{fig1}. The star markers denote the injected lens masses used in the simulations.}\label{fig2}
\end{figure*}

From the $10^4$ simulated BBH GW events, we identify eight detectable microlensing GW signals. The posterior distributions of the parameters for these events are shown in Fig.~\ref{fig1}. In the right panel of Fig.~\ref{fig1}, we compare the injected values of the GW source luminosity distance and the redshifted lens mass with their posterior estimates. 
The recovered values are consistent with the injected values within the $2\sigma$ credible regions, demonstrating the reliability of the parameter reconstruction.

\section{Hierarchical Bayesian Inference}\label{sec4}
For the compact DM population hyperparameters $\Phi$ and a set of $N_{\rm obs}$ detected microlensing BBH events, $d=\{d_1,\ldots,d_{N_{\rm obs}}\}$ with $N_{\rm obs}=8$ obtained from the $10^4$ simulated GW events in Section~\ref{sec3}, the population likelihood can be described by an in-homogeneous Poisson process. When measurement uncertainties are neglected and the selection effect is encoded in the expected number of detections $N(\Phi)$, the likelihood is given by
\begin{equation}\label{eq2-2-1}
p(d|\Phi)\propto N(\Phi)^{N_{\rm obs}}e^{-N(\Phi)}\prod_{i}^{N_{\rm obs}}p_{\rm pop}(d_{i}|\Phi).
\end{equation}
Here, $p_{\rm pop}(d_i|\Phi)$ represents the probability of observing the $i$-th event under the population model characterized by the hyperparameters $\Phi$. When the measurement uncertainty of the event-level parameters is neglected, this term can be directly described by the population distribution as Eq.~(\ref{eq2-2-1}). When measurement uncertainties are included and the selection effect is accounted for through the expected number of detections $N_{\rm det}(\Phi)$, the likelihood for a set of $N_{\rm obs}$ microlensing GW observations can be written as~\cite{Mandel2019,LVK2022,Zhou2024}
\begin{equation}\label{eq2-2-2}
\begin{split}
p(d|\Phi)\propto &N(\Phi)^{N_{\rm obs}}e^{-N_{\rm det}(\Phi)}\times\\
&\prod_{i}^{N_{\rm obs}}\int dm_{\rm L}\, {\cal L}(d_i|m_{\rm L})p_{\rm pop}(m_{\rm L}|\Phi),
\end{split}
\end{equation}
where ${\cal L}(d_i|m_{\rm L})$ denotes the single-event likelihood for the $i$-th microlensing GW event. It is obtained from the single-event parameter estimation and is related to the posterior distributions of $p(M_{\rm L}^z|d_i)$ and $p(d_{\rm L}(z_{\rm S})|d_i)$ through the their priors adopted in the single-event analysis, e.g.,
${\cal L}(d_i|M_{\rm L}^z)\propto p(M_{\rm L}^z|d_i)/\pi(M_{\rm L}^z)$. Here, $N(\Phi)$ denotes the total number of microlensing events predicted by the population model characterized by the hyperparameters $\Phi$,
\begin{equation}\label{eq2-2-3}
\begin{split}
N(\Phi)=&\int dm_{\rm L}\int dz_{\rm S}\int_0^{z_{\rm S}}dz_{\rm L} \frac{d n(m_{\rm L},\Phi)}{dm_{\rm L}}\times\\
&\frac{d\chi(z_{\rm L})}{dz_{\rm L}}(1+z_{\rm L})^2\sigma(m_{\rm L}, z_{\rm L},z_{\rm S})N_{\rm S}P_{\rm S}(z_{\rm S}),
\end{split}
\end{equation}
where $\chi(z_{\rm L})$ is the comoving distance. 
The quantities $N_{\rm S}$ and $P_{\rm S}(z_{\rm S})$ denote the number of detectable GW sources and their redshift distribution, respectively. The comoving number density of compact DM objects with an extended mass distribution $\psi(m_{\rm L},\gamma)$ is given by
\begin{equation}\label{eq2-2-4}
\frac{d n(m_{\rm L}, \Phi)}{dm_{\rm L}}=\frac{f_{\rm CO}\Omega_{\rm DM}\rho_{\rm c}}{m_{\rm L}}\psi(m_{\rm L}, \gamma),
\end{equation}
$\rho_{\rm c}$ is the critical density of the Universe. The compact DM population hyperparameters are defined as
\begin{equation}\label{eq2-2-4f}
\Phi\equiv[\gamma,f_{\rm CO}]. 
\end{equation}
The lensing cross section in Eq.~\eqref{eq2-2-3} is
\begin{equation}\label{eq2-2-5}
\sigma(m_{\rm L}, z_{\rm L}, z_{\rm S})=\frac{4\pi m_{\rm L}D_{\rm L}D_{\rm LS}}{D_{\rm S}}y_0^2.
\end{equation}
We set the maximum impact parameter to $y_0=5$, since microlensing signatures become difficult to identify for $y_0>5$~\cite{Basak2021,Zhou2022}. The normalized lens-mass distribution is then
\begin{equation}\label{eq2-2-6}
p_{\rm pop}(m_{\rm L}|\Phi)=\frac{1}{N(\Phi)}\frac{d N(m_{\rm L}, \Phi)}{dm_{\rm L}}=\psi(m_{\rm L}, \gamma).
\end{equation}
The number of detectable microlensing GW events is defined as
\begin{equation}\label{eq2-2-7}
\begin{split}
N_{\rm det}(\Phi)=&\int dm_{\rm L}\int dz_{\rm S}\int_0^{z_{\rm S}}dz_{\rm L} \frac{d n(m_{\rm L},\Phi)}{dm_{\rm L}}\times\\
&\frac{d\chi(z_{\rm L})}{dz_{\rm L}}(1+z_{\rm L})^2\sigma_{\rm det}(\lambda, m_{\rm L}, z_{\rm L},z_{\rm S})\times\\
&N_{\rm S}P_{\rm S}(z_{\rm S}),
\end{split}
\end{equation}
where $\sigma_{\rm det}(\lambda,m_{\rm L},z_{\rm L},z_{\rm S})$ is the detection cross section for microlensing signals satisfying the criterion in Eq.~\eqref{eq3-3-9}. It is given by
\begin{equation}\label{eq2-2-8}
\sigma_{\rm det}(\lambda, m_{\rm L}, z_{\rm L}, z_{\rm S})=\frac{4\pi m_{\rm L}D_{\rm L}D_{\rm LS}}{D_{\rm S}}y_{\max}(\lambda)^2,
\end{equation}
where $\lambda\in[M_{\rm L}^{z},z_{\rm S},m_1,m_2,\ldots]$ denotes the relevant set of source and lens parameters. The detected fraction of microlensing GW events is therefore defined as
\begin{equation}\label{eq2-2-8f}
\xi(\Phi)\equiv N_{\rm det}(\Phi)/N(\Phi),
\end{equation}
Finally, the posterior distribution of the population hyperparameters is obtained from
\begin{equation}\label{eq2-2-9}
p(\Phi|d)=\frac{p(d|\Phi)p(\Phi)}{Z_{\mathcal{M}}},
\end{equation}
where $p(\Phi)$ is the prior on the population hyperparameters, and $Z_{\mathcal{M}}$ is the normalization constant, or equivalently the Bayesian evidence, for the population model $\mathcal{M}$.

\section{Results}\label{sec5}
The selection effects in gravitational lensing systems and the uncertainties in the inferred lens masses are shown in Fig.~\ref{fig2}. Specifically, the left panel of Fig.~\ref{fig2} presents the detectable fraction of lensed GW events as a function of the power-law index. Our analysis reveals a positive correlation between the power-law index and the detection rate of microlensing GW events. This trend is physically expected: larger values of $\gamma$ correspond to a higher abundance of massive lenses, which typically produce stronger lensing signatures and therefore higher SNRs. In the right panel of Fig.~\ref{fig2}, we compare the inferred lens masses with the injected values used in the simulations. For each microlensing GW event, the parameter estimation provides the joint posterior samples of the source luminosity distance \(d_{\rm L}(z_{\rm S})\) and the redshifted lens mass \(M_{\rm L}^{z}\) from the MCMC chains. These samples retain the correlation between \(d_{\rm L}(z_{\rm S})\) and \(M_{\rm L}^{z}\). In the hierarchical inference, we use this two-dimensional joint posterior, rather than the product of their one-dimensional marginalized posteriors, and combine it with the optical-depth formalism in Eq.~\eqref{eq2-2-2} to derive the lens-mass likelihood \({\cal L}(d_i|m_{\rm L})\) for each microlensing GW event. The results show that the inferred lens masses are consistent with their injected values within the $2\sigma$ credible regions.

\begin{figure}
    \centering
    \includegraphics[width=0.48\textwidth, height=0.48\textwidth]{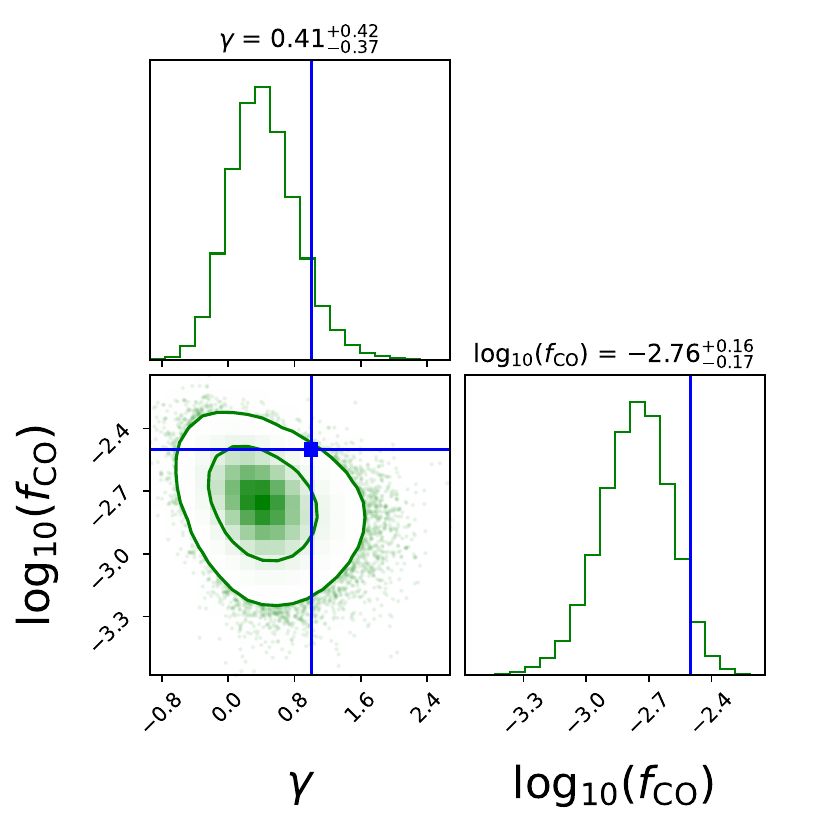}
     \caption{Posterior distributions of the population hyperparameters $[\gamma,\log_{10}(f_{\rm CO})]$, shown with the $2\sigma$ credible region. The blue solid point denotes the injected hyperparameter values used in our simulations, $[\gamma=1,\log_{10}(f_{\rm CO})=-2.5]$.}\label{fig3}
\end{figure}

For the compact DMs population hyperparameters, in the simulation, we inject $\gamma=1$ and $\log_{10}f_{\rm CO}=-2.5$. This value is not intended to represent a unique compact DMs formation scenario. Rather, it provides a simple illustrative case for testing whether the hierarchical Bayesian framework can recover the injected compact DM population parameters. We aim to reconstruct 
$\gamma$ and $f_{\rm CO}$ from the inferred lens-mass distribution. For the priors, we fix the minimum and maximum masses in the mass function to
$M_{\min}=10\,M_{\odot}$ and $M_{\max}=10^3\,M_{\odot}$, respectively. We adopt uniform priors on the power-law index and compact DMs abundance,
$\gamma\in\mathcal{U}[-1,3]$ and $\log_{10}(f_{\rm CO})\in\mathcal{U}[-4,-1]$, in Eq.~\eqref{eq2-2-9}. 
We then use the eight inferred lens masses from the microlensing GW events as input to \texttt{emcee}~\cite{emcee}, and sample the posterior distribution in Eq.~\eqref{eq2-2-9} to infer the population hyperparameters
$\Phi\equiv\{\gamma, f_{\rm CO}\}$. The resulting constraints are shown in Fig.~\ref{fig3}. We find that the best-fit values and $68\%$ credible intervals are
$\gamma=0.41^{+0.42}_{-0.37}$ and $\log_{10}(f_{\rm CO})=-2.76^{+0.16}_{-0.17}$. The blue marker in Fig.~\ref{fig3} denotes the injected hyperparameters used in our simulations, $\Phi_{\rm inj}\equiv\{\gamma=1,\log_{10}(f_{\rm CO})=-2.5\}$. Our posterior constraints are consistent with the injected values within the $2\sigma$ credible region.

\section{Conclusion}\label{sec6}
In this paper, we present a comprehensive methodology for simulating GW events lensed by compact DMs, extracting source and lens information from individual microlensing GW events, and constraining the population properties of compact DMs using an ensemble of such events. Specifically, adopting the Einstein Telescope (ET) as a representative third-generation ground-based GW detector and modeling the lenses as point masses, we infer the posterior distributions of the GW-source and lens parameters for eight detectable microlensing GW events identified among $10^4$ simulated BBH mergers. Within a hierarchical Bayesian framework, we show that, for a compact DMs population described by a power-law mass function, these eight microlensing GW events constrain the power-law index and compact DMs abundance to $\gamma=0.41^{+0.42}_{-0.37}$ and $\log_{10}(f_{\rm CO})=-2.76^{+0.16}_{-0.17}$, at the $68\%$ credible level, as shown in Fig.~\ref{fig3}. 
The inferred constraints are consistent with the injected values used in the simulations at the $2\sigma$ level. The mild offset between the recovered and injected hyperparameters is expected, since our theoretical prediction for the injected population suggests that more than ten microlensing GW events should be detectable, whereas only eight such events are identified in the simulated sample. This lower-than-expected number of detections naturally shifts the inferred constraints toward smaller values of both $f_{\rm CO}$ and $\gamma$. Our main purpose is to demonstrate that, once a small sample of microlensing GW events is available, hierarchical Bayesian inference can in principle recover information on both the compact DM abundance and the shape of its mass function.

The methodology developed in this work represents a proof-of-principle step toward using future microlensing GW detections to infer compact DMs population properties. It enables the extraction of compact DMs population information from detected microlensing signatures and provides a pathway to probing the fundamental nature of compact DMs, whether in the form of primordial black holes, exotic compact objects, or other non-particle dark-matter candidates. Although our method establishes a preliminary framework for constraining compact DMs with future GW observations, several additional effects should be incorporated to more realistically describe the underlying physical processes:
\begin{itemize}
\item 
In this work, we have focused exclusively on the point-mass lens model. However, compact DMs in the Universe may also appear in extended halo configurations, such as ``dressed PBHs''~\cite{Ricotti:2007jk,Lacki:2010zf,Adamek:2019gns,Serpico:2020ehh,Boudaud:2021irr}. If an observed signal is genuinely microlensing, analyzing it with an incorrect lens model may lead to biased parameter inference. Therefore, once a microlensing GW event is detected, it will be crucial to identify the appropriate lens model accurately~\cite{GilChoi2023,Lin2025}.

\item 
A complementary issue arises when an unlensed waveform is spuriously interpreted as a microlensing signal. For example, a GW signal from a binary on an eccentric orbit may be misidentified as microlensing if it is analyzed using quasi-circular waveform templates. In this case, lensing analyses may mistakenly attribute waveform modulations caused by orbital eccentricity to microlensing effects. Additional diagnostics are therefore required to break such degeneracies with lens parameters, such as those based on the Kramers--Kronig (KK) relation~\cite{Tanaka:2025ntr,Suyama:2025gbh}.

\item 
So far, no definitive microlensing signature has been identified in any observed gravitational-wave event. Future detections will require more complete waveform models to robustly interpret possible deviations from standard unlensed waveforms and to quantify the microlensing contribution from known stellar populations~\cite{Mishra:2021xzz,Meena:2022unp,Mishra:2023ddt,Shan:2023ngi,Shan:2024min}. Ultimately, determining whether DM substructures can be reliably distinguished from the microlensing effects of known baryonic matter remains an important open challenge.
\end{itemize}

The constraints on compact DMs are expected to improve significantly with the rapidly increasing number of microlensing GW detections in future surveys and the extended detection horizon of next-generation GW detectors. This progress will lead to substantial overlap between the parameter space probed by GW observations and that constrained by other electromagnetic observations. Consequently, combining these two powerful multi-messenger probes will enable joint constraints on both the abundance and mass distribution of compact DMs. Such joint analyses are expected to play an important role in uncovering the nature of compact DMs and their formation mechanisms, thereby providing valuable insight into the physics of the early universe.

\section{Acknowledgements}
This work is supported by the National Natural Science Foundation of China under Grants Nos.12603003, 12322301, 12275021, and 12503002; Science and Technology Research Project of Hubei Provincial Department of Education under Grants No.Q20251302; Research Performance Assesssment Grant of the Postdoctoral Fellowship Program of China Postdoctoral Science Foundation under Grant No.YJB20250367.

\bibliographystyle{apsrev4-1}
\bibliography{ref}{}

\end{document}